# Generation of 10-GHz clock sequential time-bin entanglement

Qiang Zhang[1], Carsten Langrock[1], Hiroki Takesue[2], Xiuping Xie[1], Martin Fejer[1], Yoshihisa Yamamoto[1]

[1]*Edward L. Ginzton Laboratory, Stanford University, Stanford, California 94305*
[2]*NTT Basic Research Laboratories, NTT Corporation, 3-1 Morinosato Wakamiya, Atsugi, Kanagawa 243-0198, Japan*
*qiangzh@stanford.edu*

**Abstract:** This letter reports telecom-band sequential time-bin entangled photon-pair generation at a repetition rate of 10 GHz in periodically-poled reverse-proton-exchange lithium niobate waveguides based on mode demultiplexing. With up-conversion single-photon detectors, we observed a two-photon-interference-fringe visibility of 85.32% without subtraction of accidental noise contributions, which can find broad application in quantum information and quantum entanglement research.

© 2007 Optical Society of America

---

Entangled photon pairs are regarded as the key component in quantum information and linear optics quantum computation research [1]. Especially, time-bin-type [2] entangled photon pairs at telecommunication wavelengths are promising candidates for long distance quantum key distribution (QKD) and other long distance quantum communication tasks over fiber-based networks [2] due to the low propagation loss in standard communication fiber and their insensitivity to polarization-mode dispersion compared to polarization entangled photon pairs. Therefore, there are many protocols and experiments to generate this kind of entanglement, such as parametric down-conversion processes in bulk nonlinear crystals [3] or in waveguides [4], and four-wave mixing processes in dispersion shifted fiber (DSF) [5]. However, in these experiments, the pump pulse repetition rate was limited by the performance of the single-photon detectors and was on the order of 100 MHz [3-5].

There have been few quantum-communication experiments at gigahertz clock rates and, very recently, some authors of this letter have developed a correlated photon-pair system at a 10-GHz clock rate [6]. Here, we present 10-GHz-clock sequential time-bin entangled photon pair generation in reverse-proton-exchange (RPE) periodically poled lithium niobate (PPLN) waveguides based on mode demultiplexing and parametric down-conversion. Up-conversion-assisted fast silicon avalanche photo diodes (APDs) with a timing jitter of 45 ps (FWHM) and a dark count rate of 3E-6 per time window are used to count the entangled photons in the telecom band. RPE waveguides are used due to their large nonlinear efficiencies and low propagation loss (~0.1 dB/cm) [7]. A two-mode interaction with mode demultiplexing using asymmetric Y-junctions is used to generate and to separate the degenerate co-polarized photons on chip [6,7]. In our coincidence measurement, a two-photon-interference-fringe visibility of 85.32% is achieved with an entangled-photon-pair flux of 313 Hz. All the results are achieved without subtracting any accidental noise.

The basic idea of time-bin entanglement is to use two consecutive laser pulses to pump a crystal possessing a $\chi^{(2)}$ nonlinearity. The time interval between the two pulses is $\tau_1$, and the pulse duration is $\tau_2$ ($\tau_2 \ll \tau_1$). The coherence time of the laser pulse is much longer than the time interval $\tau_1$ [8] so that the phase between two consecutive pulses is stable, in our case, defined as $\phi_p$. When the two pulses pass through the nonlinear crystal, both have a certain probability to generate a photon pair via the process of parametric down-conversion. If there is no other way except for the time difference to distinguish the two possible generations, the quantum state of the photon pair is given by, $\left(|t_1\rangle|t_1\rangle + e^{j\phi_p}|t_2\rangle|t_2\rangle\right)/\sqrt{2}$, where $t_1$ and $t_2$ ($t_2 - t_1 = \tau_1$) represent the generating time of the photon pair. This type of entanglement has been widely used in QKD and quantum teleportation while people are now trying sequential time-bin entanglement [9].

In the sequential time-bin condition, not two laser pulses but a laser pulse train is being used. Each pulse duration is $\tau_2$ with time interval $\tau_1$, while the phase between each consecutive pulse in the train is $\omega_p * \tau_1$, where $\omega_p$ denotes the angular frequency of the laser. Therefore, after pumping the nonlinear crystal, the quantum state is,

$$|t_1\rangle_s|t_1\rangle_i + e^{j\omega_p\tau_1}|t_2\rangle_s|t_2\rangle_i + \cdots + e^{jn\omega_p\tau_n}|t_n\rangle_s|t_n\rangle_i + \cdots \quad (1),$$

where s and i represents signal and idler photon.

Our entanglement is a sequential time-bin entanglement at a 10-GHz clock rate. Instead of using mode-locked laser pulses as described in reference [6,9], we use a Mach-Zehnder (MZ) interferometer to modulate a CW laser with a wavelength of 1559nm at 10 GHz, increasing the phase stability. In order to obtain shorter pulse, the bias voltage of the modulator was set at a small value and the modulator was operated in nonlinear regime. The driving RF signal is derived from a 10-GHz synthesizer resulting in modulated pulse durations of approximately 40 ps and 100 ps time intervals (i.e. $\tau_1 = 100\,ps$, $\tau_2 = 40\,ps$) as shown in Fig. 1.

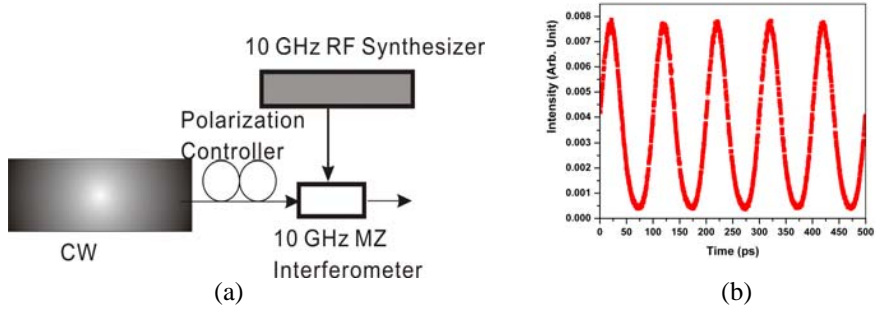

(a)           (b)

Fig. 1 Modulated 10-GHz laser-pulse setup (a) and pulse train (b). An in-line polarization controller is used to change the polarization meeting the polarization dependence of the interferometer. The pulse-train picture is captured from a 40-GHz oscilloscope.

In our experiment, the 10-GHz laser pulse train first passes through a RPE PPLN waveguide to generate a second-harmonic pulse at 779.5 nm to serve as the pump of parametric down-conversion in a second waveguide device. As shown in Fig. 2, the residual 1559 nm light is strongly attenuated with four dichroic mirrors before launching the 779.5 nm pump into the second RPE PPLN waveguide containing an asymmetric Y-junction mode multiplexer/demultiplexer and quasi-phase-matching gratings. The Y-junction PPLN waveguide has been used in an earlier experiment to generate correlated photon pairs [6]. The generated degenerate entangled photon pairs have a 40-nm bandwidth centered around 1559nm.

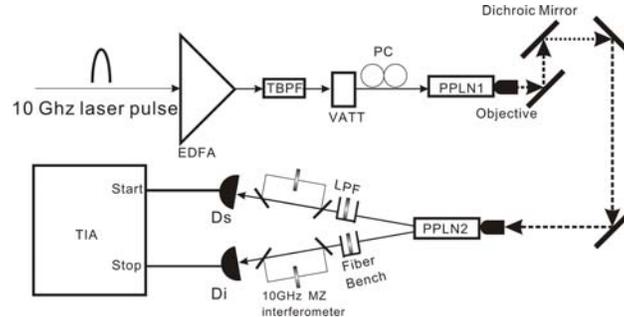

Figure 2. Setup for 10-GHz sequential time-bin entanglement generation. A 10-GHz laser pulse train is amplified by an erbium-doped fiber amplifier (EDFA) and spontaneous-emission generated by the EDFA is cut by a tunable band-pass filter (TBPF) with a 1-nm bandwidth. A variable attenuator (VATT) is used to control the laser power and a polarization controller is used to launch the proper polarization into the PPLN waveguide. PPLN1 is the waveguide used for second-harmonic generation (SHG) of 779.5-nm pump pulses and PPLN2 is the Y-junction waveguide used for parametric down-conversion. Between them are four dichroic mirrors to greatly reduce the residual 1559-nm photons. The two outputs of PPLN2 are fiber pigtailed and two fiber U-benches both with a long pass filter in them are utilized to eliminate the 779.5 nm pump. $D_s$ and $D_i$ are two up-conversion-assisted Si APDs.

To characterize the generated photon pairs, the signal and idler photons are coupled into two-unbalanced planar lightwave circuit (PLC) MZ interferometers to measure the interference-fringe visibility [5]. The MZ interferometer is a temperature controlled silica-on-silicon PLC and unbalanced by 100 ps, the same as the time interval of the pump laser. Therefore, the interferometer can convert a state $\left|t_x\right\rangle_y$ into ($\left|t_x\right\rangle_y + e^{i\theta_y}\left|t_x\right\rangle_y$), where $t_x$ represents the photon-pair generation time and $\theta_y$ is the phase difference between the two paths of the interferometer for mode y, which can be changed by adjusting the temperature of the interferometer. Then the output state from the MZ interferometer will be,

$$|t_1\rangle_s|t_1\rangle_i + (e^{jw_p\tau_1} + e^{j(\theta_s+\theta_i)})|t_2\rangle_s|t_2\rangle_i + \cdots + e^{j(n-1)w_p\tau_n}(e^{jw_p\tau_1} + e^{j(\theta_s+\theta_i)})|t_n\rangle_s|t_n\rangle_i + \cdots \quad (2),$$

where a normalization constant is omitted and noncoincident terms are discarded since we only post-select the coincidence events in the experiment. Thus we can observe a two-photon interference at all time bins except for the first and last bin.

The outputs of the interferometers are detected by fast single-photon detectors $D_s$ and $D_i$, connected to a time-interval analyzer (TIA) to generate start and stop signals, respectively, for the coincidence measurement.

In such a high-speed experiment, the single-photon detector's timing jitter is a crucial parameter. One must choose a single-photon detector with a timing jitter less than the time interval of the laser pulse, e.g. 100 ps in our 10-GHz-clock system. While, we used a superconductor single-photon detector to observe 10-GHz correlated photon pairs in a previous experiment [6], here we use up-conversion-assisted Si APDs. The Si APDs used here (id Quantique id100-20) have a 40-60 ps timing jitter (FWHM) in the spectral range from 400 to 900 nm and a 15% quantum efficiency (QE) at 700 nm. We equipped two of these APDs with our PPLN-based up-conversion device, which can up-convert a 1.55-μm photon to a 713-nm photon using a strong 1.32-μm pump laser [10]. Our up-conversion single photon detector has a 0.2-nm wide 3 dB acceptance bandwidth [10], which is much narrower than the 40-nm bandwidth of the entangled photon pairs. Therefore, only a small fraction of the photon pairs can be detected, which effectively reduces the QE of the detection system to 2% using a 100-mW pump power. The dark count rate at this pump power is about 40 kHz.

To determine the timing jitter during the coincidence measurement, we performed a measurement of the photon-pair-correlation timing jitter. The setup is similar to the entanglement setup in Fig. 2 except that we input 1-GHz pump laser pulses (10 ps pulse duration) to generate correlated photon pairs instead of 10 GHz and that the photon pairs are directly detected by our up-conversion single-photon detectors without passing through the PLC interferometers. The detection signal from $D_s$ will be input to the start channel of the TIA, while $D_i$'s signal will be sent to the stop channel. A histogram obtained by the TIA shows the time-correlation spectrum of the photon pairs as shown in Fig. 3. Since the two detection signals are from correlated pairs and the pump laser's pulse duration is much smaller compared to the timing jitter of the Si APD, the uncertainty of the time histogram is mainly due to the combined timing jitter of the two up-conversion single-photon detectors. From the histogram, we can estimate an 80-ps timing jitter (FWHM) and 200-ps full width at tenth maximum (FWTM) for the coincidence events.

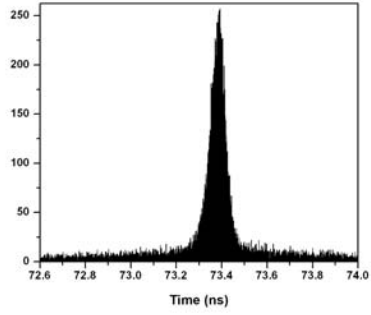

Figure 3. Histogram of time spectrum of the correlated photon pairs.

In the experiment, we launch 2 mW pump power into the Y-junction PPLN waveguide and measured the coincidence to accidental coincidence ratio (CAR) [5,6]. The CAR is a good figure of merit to demonstrate the contribution from the noise in the coincidence measurement and also provides an estimate of the average photon-pair number and the entanglement visibility. The setup to measure the CAR is similar to the setup in Fig. 2 without the two PLC MZ interferometers. Then we implemented a start-stop measurement for the photon pairs and the detection results of the signal and idler photons are input to the start and stop channel of TIA, respectively [5,6]. A coincidence in the matched time slot is the true coincidence caused by photons generated with the same pump pulse and a coincidence in the unmatched time slot is the accidental coincidence caused by photons generated by different pump pulses. The ratio between the two coincidences is the CAR. In our experiment, the CAR reaches 26 with 2 mW pump power when the average number of photon pairs per pulse is around 0.03. With this CAR, we estimate the visibility of the entanglement to be 94% [5].

To observe the interference pattern, we set the phase $\theta_s$ in the signal channel and varied the phase $\theta_i$ in the idler channel by adjusting the temperature of the two interferometers. We first set the temperature of the MZ interferometer in the signal channel to 25.71 °C, and varied the temperature of the other interferometer in the idler channel. We achieve an interference pattern with a visibility of $(87.19 \pm 5.78)\%$. To demonstrate entanglement, one interference pattern is not sufficient; at least one other pattern in a non-orthogonal basis is necessary. To observe this pattern, we changed the temperature of the signal interferometer to 27.51 °C and observed another interference-fringe pattern with a visibility $(83.44 \pm 5.76)\%$ as shown in Fig. 4. The two curves with an average visibility of $(85.32 \pm 5.77)\%$, which is well beyond the visibility of 71% necessary for violation of the Bell inequality [11], demonstrate entanglement. The imperfection of the visibility compared to the 94% estimated visibility is mainly caused by the 200-ps-FWTM timing jitter and the 1% base-line error of the PLC MZ interferometer.

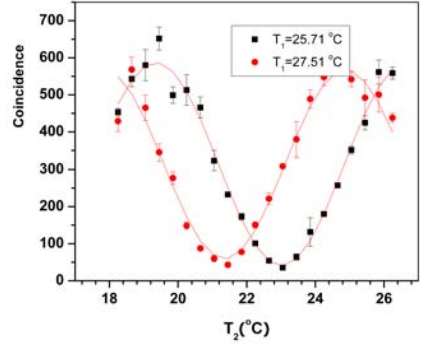

Figure 4. Coincidence interference fringes observed in the experiment. T1 and T2 are the temperatures of the PLC interferometers in the signal and idler channels, respectively. All coincidence points in the figure are derived from a one million start triggers for the TIA.

The total system loss is 26 dB and each item's contribution can be seen in Table. 1. Including all the loss terms and the 2% quantum efficiency of our up-conversion detector, we achieved a 313 Hz entangled photon-pair flux.

Table 1. Loss distribution of the whole system

| Item | Fiber pigtailing and propagation loss | Fiber U-bench and filtering loss | Insertion loss of PLC MZI |
|---|---|---|---|
| Loss (dB) | 10 dB | 11 dB | 5 dB |

To summarize, we experimentally generated 10-GHz-clock sequential time-bin entanglement in the telecom band with a RPE PPLN waveguide. The visibility of the achieved entanglement is 85.32% without subtraction of any noise terms, with a photon pair flux of 313 Hz. If we only consider the entanglement source itself and assume 100% efficient single-photon detectors, our setup would generate 780 kHz photon pair with the same visibility. Improving the fiber pigtailing of the waveguide can make the source brighter by 7 dB, i. e. 4 MHz, which can find immediate application in long distance quantum key distribution and other quantum communication protocols.

**Acknowledgement**

The authors thank N. Gisin, R. Thew and Id Quantique Inc. for lending two id100-20 Si APDs and thank L. G. Kazovsky, W. Tao for lending the 10-GHz EO modulator. This research was supported by the MURI center for photonic quantum information systems (ARO/ARDA program DAAD19-03-1-0199), SORST, CREST programs, Science and Technology Agency of Japan (JST), the U.S. Air Force Office of Scientific Research through contracts F49620-02-1-0240, the Disruptive Technology Office (DTO). We acknowledge the support of Crystal Technology, Inc.